# TDD for Embedded Systems: A Basic Approach and Toolset

Rogerio Atem de Carvalho, Hudson Silva, Rafael Ferreira Toledo, Milena Silveira de Azevedo

Scientific Computing Group (C2), Centre for Embedded and Aerospace Systems (CRSEA), Instituto Federal Fluminense (IFF)
Campos dos Goytacazes, Brasil
`ratem@iff.edu.br, silvaferreira.hsf@gmail.com, rafaelt91@gmail.com, milena.slvr@gmail.com`

**Abstract.** The evolution of information technology and electronics in general has been consistently increasing the use of embedded systems. While hardware development for these systems is already consistent, software development for embedded systems still lacks a consolidated methodology. This short paper describes a process and toolset for Embedded Systems Modeling and Verification using FSM (Finite State Machines) and TDD (Test-Driven Development).

## 1 Introduction

The evolution of information technology and electronics in general has been consistently increasing the use of embedded systems. Those systems can be defined as computing devices with a specific purpose, usually integrated to an external system, which perform some specific function. Nowadays, embedded systems have several applications such as traffic management, car navigation, industrial control etc. An embedded system has the following characteristics [1]:

•   Its functionalities and performances are deeply dependent of mechanic, electronic, hardware and software technologies elements integration.

•   Its hardware resources, such as memory size and processing power, are usually limited.

•    It has to be robust, thus its behavior must be tightly controlled even when the systems have failures.

The specification of systemic requirements is a complex step and should be verified several times during the project lifetime cycle. To facilitate this step, notations that represent business process obtain these requirements. In this sense, FSM is the notation widely chosen for modeling  embedded systems.

 A FSM models a process by identifying which states the system can be in, which inputs or events trigger the state transitions, and how the system will behave in each state. In this model, the software execution is viewed  as a sequence of transitions that move the system through its various states. [2]

The aim of this paper is to present a toolset and method to facilitate the development of the software part of embedded systems. This tooling concerns coding as well as systemic validation and modeling. In order to reach such goal, this paper is briefly present a toolset and a proposed process, followed by some conclusions and some current and future research directions on the development of embedded systems.

## 2 Toolset

The toolset is based on open source tools, both developed by communities, such as Yakindu and Google C++ Testing Framework, and developed by some of the authors, such as Clover and Test Suite.

### 2.1 Yakindu

Yakindu Statechart Tools (SCT) consists of an integrated modeling environment for the specification and development of reactive, event-driven systems using the concept of Finite State Machines (FSM). This user-friendly tool includes sophisticated graphic edition, validation and simulation, as well as code generator in Java, C and C++ [3]. Yakindu was used as an Eclipse (a Java based software development platform) plugin for generating tests skeletons, integrating with a test harness and code executing during simulation.

### 2.2 Clover

One of the basic TDD's techniques is the use of Test Doubles. When a test is written and it is not possible, or not desirable, to use a real dependency, it can be substituted by a double [4]. However, embedded systems are usually implemented in the C programming language, which is not an object oriented one, therefore, in this case, doubles are used to substitute not objects, but function calls. In order to achieve this, the authors developed a library called C Libraries Overloading (Clover). Clover is a powerful tool with the following capabilities:
- Any C function can be overloaded, providing that a double implementation is supplied to it.
- Clover is a Dynamic Library, therefore, it is not necessary to introduce any change in the production code (Unit Under Test – UUT), it is linked together with the UUT only during the tests.
- Functions calls can be doubled in various situations: during the execution of a given piece of code, during all the time, or in a given call. In that way. In that way function calls to functions that are used to implement other library functions, such as malloc() or fprintf(), suffer from no interference and work normally.

Figure 1 shows an example use of Clover, testing malloc(), set_status() is used to intercept malloc() calls and detour them to the double behaviour.

```c
void it_overrides_malloc(void){
    void* pt;

    pt = allocate(10);
    assert( pt != NULL );
    free(pt);
    set_status(_malloc, 1, 0, NULL);
    pt = allocate(10);
    assert( pt == NULL );
    //Checks deactivation
    pt = allocate(10);
    assert( pt != NULL );
    free(pt);
    //Always activated
    set_status(_malloc, -1, 0, NULL);
    pt = allocate(10);
    assert( pt == NULL );
    set_status(_malloc, 0, 0, NULL);
}
```

**Figure 1**. Testing malloc() using Clover.

Clover is a vital element of the proposed testing suite, since it can simulate in a very cheap and automated way failures such as lack of memory, error in file reading or writing, or any kind of error behavior of any C library function, covering all the possible errors that a function library call can return, which is highly valuable to critical systems.

## 2.3 Test Suite Plugin

After modeling the FSM, or parts of it in fact, the automated tests skeletons are generated with the Test Suite module, which is also an Eclipse plugin. This module asks for three sets of information to the user, as shown in Figure 2.

**Figure 2**. Test Suite GUI

The Expectations field should include all the expected destiny states according to the FSM model. Variables are responsible for defining the sequence of variables, which are verified during events transitions. Finally, Inputs field receives the sequence of values, which the variables need to trigger the state transition.

Currently, the testing harness in use is the Google C++ Testing Framework, which offers an easy integration with Eclipse IDE. Google Test supports automatic test discovery, a rich set of assertions, user-defined assertions, value- and type-parameterized tests, among other features [5]. However, it is envisioned the use of Test High C (THC), a lightweight, and easier to extend test harness.

## 3  Proposed Process

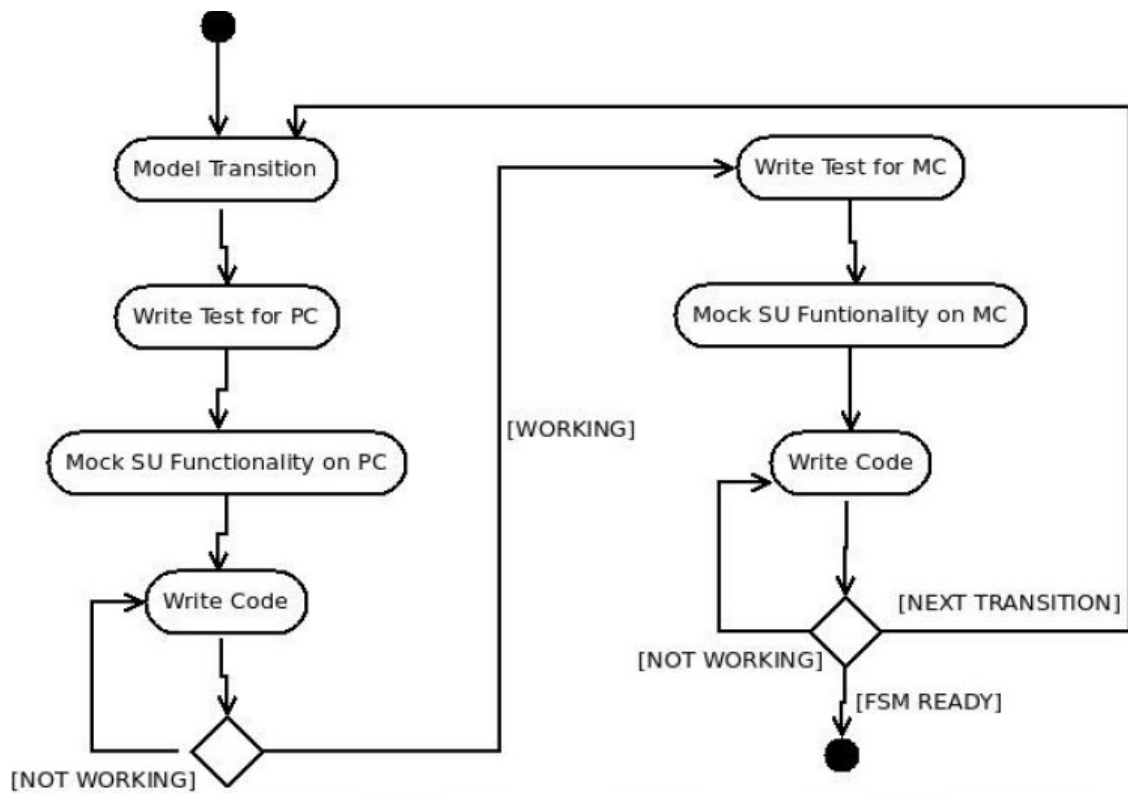

**Figure 3**. Proposed Development Process.

Figure 3 presents the proposed process for developing embedded systems using TDD. Firstly, the development flow starts with the modeling of a single transition of the FSM that represents the embedded system, then a single test for this transition is created, using a PC architecture to run it. Still using a standard PC hardware, the next step is to mock the hardware (SU[1]) behavior through software. Finally, the hardware behavior is mocked on a microcontroller, with the aim of guaranteeing a closer simulation of the hardware. In that way, it is possible to develop hardware and software in parallel, or the second before the first is done.

---

1  SU stands for Science Unit, which is satellite's payload hardware used as the first case study of the proposed methodology.

### 3.1 Simulation

Besides modeling, Yakindu enables the simulation of the system behavior, using the FSM notation. Figure 4 shows the simulation of a model being executed by Yakindu. In this modeling example, there are three states (State1, State2, and State3). In addition, each state expects to receive some information. If the information is in accordance with the transition conditions, this transition state is triggered. The next state to be activated is indicated by the arrows. In the figure above, the current state is State2. The simulation, which needs no coding, helps the user to identify failures in the specification of systemic requirements since the main idea of the model can be validated and simulated visually.

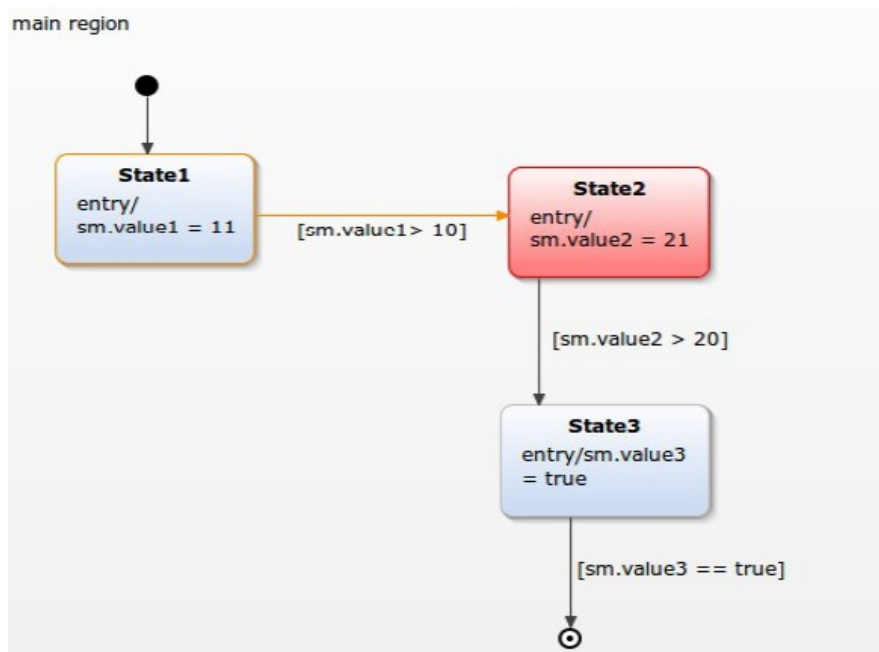

**Figure 4**. Yakindu example simulation.

### 3.2 Test Code Generation

With the information provided by the user, the Test Suite module generates code, responsible for running the FSM simulation and for checking the assertions of the desired FSM behavior. Figure 5 presents and example file with the automated tests generated.

```cpp
1  #include <stdio.h>
2  #include <stdlib.h>
3  #include "src-gen/sc_types.h"
4  #include "src-gen/Sm.h"
5  #include "testinglib/gtest/gtest.h"
6
7  class TestStatemachine: public ::testing::Test {
8      protected:
9      Sm handle;
10     SmStates sm_main_region_State1;
11     SmStates sm_main_region_State2;
12     SmStates sm_main_region_State3;
13     SmStates sm_main_region__final_;
14 };
15
16 TEST_F(TestStatemachine, testsm) {
17     sm_init(&handle);
18     sm_enter(&handle);
19
20     EXPECT_TRUE(sm_isActive(&handle, sm_main_region_State1));
21
22     smIfaceSm_set_value1(&handle, 13);
23     EXPECT_TRUE(sm_isActive(&handle, sm_main_region_State2));
24
25     smIfaceSm_set_value2(&handle, 54);
26     EXPECT_TRUE(sm_isActive(&handle, sm_main_region_State3));
27
28     smIfaceSm_set_value3(&handle, true);
29     EXPECT_TRUE(sm_isActive(&handle, sm_main_region__final_));
30 }
```

**Figure 5**. TestStateMachine.cpp File.

The assertion EXPECT_TRUE verifies if the function input is equal to true. The set of commands contained in the file consists of the repeated process of origin state verification, transition conditions execution and destiny state verification until the user inputs have finished.

## 4 Conclusions

This paper presented very briefly a methodology and a supportive toolset for developing embedded systems software with the use of TDD. Yakindu FSM modeling and simulation tool is used for requirements validation while tools developed and integrated by the authors are used to support the verification part of the process.

The resources here presented are being employed in the development of the Brazilian 14-BISat nanosatellite, specifically on the development of the embedded software that is responsible for controlling its payloads. This satellite is part of the international QB50 mission, and will perform the function of server for a network of satellites.

Currently, the Test Suite tool is being modified in order to manipulate more granular constructs, in others words, to fully support the iterative and incremental development process, and to integrate Behavior Driven Development (BDD) techniques as development resources.

# 5 Referencies